\documentclass{pasj00}

\begin{document}
\SetRunningHead{Grechnev et al.}{Microwave bursts and proton
events}

\title{Relations between strong high-frequency microwave bursts and proton events}

\author{Victor \textsc{Grechnev}\altaffilmark{1},
Natalya \textsc{Meshalkina}\altaffilmark{1}, Ilya
\textsc{Chertok}\altaffilmark{2}, Valentin
\textsc{Kiselev}\altaffilmark{1}}
 \altaffiltext{1}{Institute of
Solar-Terrestrial Physics SB RAS, Lermontov St. 126A, Irkutsk
664033, Russia}\email{grechnev@iszf.irk.ru}
 \altaffiltext{2}{Pushkov Institute of Terrestrial Magnetism,
Ionosphere and Radio Wave Propagation (IZMIRAN), Moscow, 142190
Russia}\email{ichertok@izmiran.ru}


\maketitle

\begin{abstract}

Proceeding from close association between solar eruptions, flares,
shock waves, and CMEs, we analyze relations between bursts at 35
GHz recorded with the Nobeyama Radio Polarimeters during
1990--2012, on the one hand, and solar energetic particle (SEP)
events, on the other hand. Most west to moderately east solar
events with strong bursts at 35 GHz produced near-Earth proton
enhancements of $J(E > 100 \ \mathrm{MeV}) > 1$ pfu. The strongest
and hardest those caused ground level enhancements. There is a
general, although scattered, correspondence between proton
enhancements and peak fluxes at 35 GHz, especially pronounced if
the 35 GHz flux exceeds $10^4$ sfu and the microwave peak
frequency is high. These properties indicate emission from
numerous high-energy electrons in very strong magnetic fields
suggesting a high rate of energy release in the flare--CME
formation process. Flaring above the sunspot umbrae appears to be
typical of such events. Irrespective of the origin of SEPs, these
circumstances demonstrate significant diagnostic potential of
high-frequency microwave bursts and sunspot-associated flares for
space weather forecasting. Strong prolonged bursts at 35 GHz
promptly alert to hazardous SEP events with hard spectra. A few
exceptional events with moderate bursts at 35 GHz and strong
proton fluxes look challenging and should be investigated.

\end{abstract}

\section{Introduction}
 \label{s-introduction}

Solar energetic particles (SEP), which are somehow accelerated in
association with solar eruptive events, offer hazards for
equipment and astronauts on spacecraft and even for passengers and
crew members on aircraft in high-latitude flights due to secondary
particles. The highest-energy extremity of SEP events sometimes
produces in the Earth's atmosphere considerable fluxes of
secondary neutrons, which are able to cause ground-level
enhancements (GLE) of cosmic ray intensity registered
preferentially with high-latitude neutron monitors (see, e.g.,
\cite{Cliver2006,Aschwanden2012,Nitta2012}). The lowest latitude,
at which a GLE is observed, is determined by the energy of primary
particles.

Existing methods to diagnose SEP productivity of a solar eruption,
which just occurred, are not yet perfect. Still larger
uncertainties exist in forecasting SEP events. Elaboration of
existing methods call for better understanding of particle
acceleration, when and where it occurs, and in which conditions.

SEPs mainly consist of protons, alpha particles, and heavier ions.
Their energies range from tens to hundreds of MeV, and sometimes
up to several GeV. Unlike electrons, which are widely manifest in
all layers of the solar atmosphere practically in the whole
observable range of electromagnetic emissions, starting from
gamma-ray bremsstrahlung continuum and up to metric radio waves,
energetic protons on the Sun can only be detected from gamma-ray
emissions appearing in their interactions with dense material
(see, e.g., \cite{VilmerMacKinnonHurford2011}). These are discrete
gamma-ray lines in a range of 0.5--10 MeV produced by nuclei with
energies of a few tens of MeV, and a very broad line around 70 MeV
produced in the decay of neutral pions, which appear in
interactions of protons with energies exceeding 300~MeV. The
$\pi^{0}$-decay emission can only be identified with
high-sensitivity gamma-ray spectrometers in big flares, and
therefore the number of all events, in which this emission has
been detected so far, starting from its first observation reported
by \citet{Forrest1986}, is as small as one dozen (see
\cite{ChuppRyan2009,Kurt2013,VilmerMacKinnonHurford2011} for the
review; recent case studies, e.g., by
\cite{Grechnev2008,Kuznetsov2011,Ackermann2012}). Imaging in
nuclear gamma-ray lines is only possible from data of Reuven
Ramaty High-Energy Solar Spectroscopic Imager (RHESSI,
\cite{Lin2002}) and does not exceed the energy of the 2.22 MeV
line. One more source of information about accelerated particles
on the Sun is provided by solar neutrons, which are produced in
collisions of high-energy protons, and are sometimes registered
with low-latitude neutron monitors on the sunlit side of Earth
(e.g., \cite{Watanabe2003,VilmerMacKinnonHurford2011}) as well as
some space-borne detectors.

It is possible to follow propagation in the interplanetary space
of energetic electrons from their signatures in decameter to
kilometer radio waves. On the other hand, the lack of
observations, which could track heavy energetic particles from the
Sun to Earth, hampers understanding their origin. There are two
different major viewpoints on the origins of SEPs in the
interplanetary space (see, e.g.,
\cite{Kallenrode2003,Grechnev2008,Reames2013} and references
therein). One concept relates SEPs with flare processes within an
active region (e.g., \cite{KleinTrottet2001,Aschwanden2012}).
According to the different concept, SEPs are accelerated to high
energies by a bow shock driven by the outer surface of a
super-Alfv{\'e}nic CME (e.g., \cite{Cliver1982};
\authorcite{Reames1999}
\yearcite{Reames1999,Reames2009,Reames2013};
\cite{Gopalswamy2012}). The seemingly incompatibility of the two
concepts is due to the traditional idea that particle acceleration
occurs either (i)~within a closed flaring solar active region or
(ii)~rather far from the Sun by a shock wave, whose properties are
determined by the CME and not by the flare. The CME and flare are
considered to be independent of each other.

On the other hand, studies of last years, supported by the
increasing observational material, specify long-standing issues
and update corresponding concepts. The CME acceleration turns out
to be closely associated with a flare and occur simultaneously
with HXR and microwave bursts (\cite{Zhang2001};
\authorcite{Temmer08} \yearcite{Temmer08,Temmer10};
\cite{Grechnev2011_I}). The helical component of the CME's
magnetic flux rope responsible for its acceleration is formed by
reconnection, which is also responsible for a flare
\citep{Qiu2007,Temmer10}. There is a detailed quantitative
correspondence between the reconnected flux and the rate of energy
release during a flare (e.g., \cite{Miklenic2009}). Thus,
parameters of a CME should correlate with parameters of the
associated flare, that has been established indeed (e.g.,
\cite{Vrsnak2005}).

Furthermore, \authorcite{Grechnev2011_I}
(\yearcite{Grechnev2011_I,Grechnev2011_III}) and
\citet{AfanasyevUralov2011} have shown that shock waves, most
likely, are excited by erupting flux ropes as impulsive pistons
inside developing CMEs, and this occurs during hard X-ray and
microwave flare bursts (i.e., at the rise phase of the soft X-ray
emission). Then, the shock wave detaches from the piston, and
quasi-freely propagates afterwards like a decelerating blast wave.
The transition of the shock wave to the bow-shock regime is
expected later, probably, beyond the field of view of LASCO/C3. As
the results of \citet{Reames2009} show, the release of SEPs near
the Sun occurs within a few solar radii. Parameters of the shocks
at such distances should be mainly determined by their initial
sources and therefore related to parameters of the associated
flares. Note also that from the preceding paragraph it follows
that the early evolution of the CME speed is expected to be
roughly proportional to the soft X-ray flux time profile at the
rising phase. That is, if the soft X-ray flux increases gradually,
then the development of a shock wave is not expected at small
distances from the Sun.

CMEs should favor escape of flare-accelerated particles trapped in
the flux rope (K.-L.~Klein, 2011, private communication).
Reconnection of the expanding CME's flux rope with a coronal
streamer allows trapped particles to access magnetic fields, which
are open into the interplanetary space, and facilitates their
escape, thus solving the problem discussed by \citet{Cliver1989}.
A possible extremity is presented by jet-like eruptions, when
reconnection decomposes the magnetic flux rope (e.g.,
\cite{Meshalkina2009}).

These considerations show that the traditional contrasting the
flare-acceleration and shock-acceleration options is probably
exaggerated. There are convincing arguments in favor of either
option. While gamma-rays nearly concurrent with different flare
emissions favor flare-acceleration of heavy particles
simultaneously with electrons, there is no reason to object
shock-acceleration, if in situ measurements of the SEP composition
such as the iron charge state, Fe/O ratio, and other parameters
indicate acceleration of ions at normal coronal temperatures (see,
e.g.,
\authorcite{Reames2009} \yearcite{Reames2009,Reames2013}). On the
other hand, in situ measurements are limited to moderate energies
of ions, while acceleration of heavier particles is more efficient
by Fermi processes operating in shock-acceleration indeed. It is
possible that the contesting concepts are based on different
observations subjected to selection effects.

For all these reasons, it seems to be logical to expect the
correspondence between parameters of SEPs and microwave bursts.
Indeed, the correlation between SEP events and strong
high-frequency radio bursts has been known for many decades (e.g.,
\cite{Croom71}). On the other hand, \citet{Kahler82} explained
this association by the `Big Flare Syndrome', i.e., a general
correspondence between the energy release in an eruptive flare and
its various manifestations, whereas the actual SEP acceleration
was considered by a CME-driven bow shock. Later exaggerations of
the shock-acceleration concept have led to underestimation of
diagnostic opportunities of microwave bursts. However, it seems
worth to analyze relations between flare microwave bursts and SEPs
irrespective of their origins. Some aspects of the correspondence
between parameters of flares, CMEs, shock waves, and SEPs have
been really stated by \citet{NittaCliverTylka2003} and
\citet{Gopalswamy2012}.

Relations between SEP events and microwave bursts at 9 GHz were
considered by \citet{Akinian1978,Cliver1989}. This frequency can
belong to either optically thin or thick branch of the
gyrosynchrotron spectrum that complicates the relation.
Higher-frequency emissions in the optically thin regime seem to be
most sensitive to large numbers of high-energy electrons gyrating
in strong magnetic fields, being thus directly related to the rate
of energy release in the flare--CME formation process. The
frequency of 35~GHz is the highest one, at which stable long-term
observations are available thanks to the operation of the Nobeyama
Radio Polarimeters (NoRP, \cite{Nakajima1985}). In the present
study, we analyze relations between microwave bursts recorded with
NoRP since 1990, on the one hand, and large high-energy proton
enhancements, on the other hand. Systematic lists are available of
data on proton events (e.g., \cite{Kurt2004,Chertok09}) and
especially on GLE events (e.g., \cite{Cliver2006}).

\citet{Chertok09} found that the $F_{35} > 10^4$ sfu criterion
selected SEP-productive events, and not only west ones. We extend
our analysis to a larger set of events with $F_{35} > 10^3$ sfu
occurring during the NoRP observational daytime. To reveal the
events missed by our criterion, we also consider all near-Earth
$>100$ MeV proton enhancements exceeding $J_{100} > 10$ pfu [1 pfu
= 1 particle/(cm$^{2}$~s$^{-1}$~ster$^{-1}$)], whose solar source
events could be observed in Nobeyama.

The input data, their processing, and the output parameters are
considered in section~\ref{s-data}. Section~\ref{s-results}
analyzes the parameters of near-Earth proton enhancements versus
parameters of microwave bursts estimated in Section~\ref{s-data}.
Besides the proton-rich events associated with intense bursts at
35 GHz, seven additional events have been revealed from the lists
of SEP events by the $J_{100} > 10$ pfu criterion. Three of these
SEP enhancements were due to solar backside events, whose
microwave emissions were occulted. The 35~GHz fluxes for four
remaining exceptional proton-rich events ranged from 140 to 780
sfu. Section~\ref{s-discussion} shows that a proton-productive
event can be expected if a related flare occurs in strong magnetic
fields, especially in those associated with sunspots; discusses
possible reasons for the exceptional events; and proposes a
tentative relation between the peak flux observed at 35 GHz and an
expected importance of a SEP event. Section~\ref{s-conclusion}
briefly summarizes the outcome of our analysis.

\section{Data}
 \label{s-data}

\subsection{Processing of NoRP Records}

Data on solar microwave bursts recorded with NoRP are
automatically processed by software and posted on the NoRP Web
site. The results are sometimes insufficiently accurate, e.g., in
evaluation of the pre-burst level, and can suffer from various
problems. These circumstances required examination of all records
in question and evaluation of parameters in the interactive mode.

NoRP records at 35 GHz for some events were absent or damaged. In
such cases, the value of $F_{35}$ was estimated by means of
interpolation from the 17 GHz and 80 GHz data. The 80 GHz fluxes
during 1995--2005 were corrected with a time-dependent factor of
$[T_{\mathrm{year}}/1995.83]^{630}$ (H.~Nakajima 2005, private
communication). The NoRP did not operate at both 35 GHz and 80 GHz
on 2004 November 10; the parameters of this event were roughly
estimated from lower-frequency data and correlation plots of the
Nobeyama Radioheliograph (NoRH; \cite{Nakajima1994}). Similarly we
had to deal with the 2002 April 21 event.\footnote{These two
events are shown in Figure~\ref{fig:event_distribution} with the
filled squares.}

The turnover frequency $f_\mathrm{peak}$ of the microwave spectrum
was computed from NoRP total flux data by using the parabolic fit of
the averaged log--log spectrum near the peak of the burst. This way
was previously used by \citet{White2003} and \citet{Grechnev2008}.
Here we used data of lower-frequency polarimeters, which were
located before 1994 in Toyokawa \citep{Torii1979}.

The time profiles of the bursts at 35 GHz are sometimes complex,
contain more than one peak, etc. They are typically shorter and
more impulsive than those at lower microwaves and quite different
from soft X-ray ones, whose time profiles are intrinsically
gradual. It is difficult to measure the duration of a burst at 35
GHz, $\Delta t_{35}$, with a simple formal criterion, e.g., by
referring to 0.5 or to 0.1 \& 0.9 levels. We have not yet found an
adequate formal criterion to characterize durations of complex
bursts, and therefore estimated $\Delta t_{35}$ manually by
marking characteristic durations. These estimates are rather
subjective and need elaboration. Nevertheless, they represent a
general tendency correctly. The tendency remains, if the formal
0.5-level criterion is used, while the durations, $\Delta t_{35}$,
become shorter.

\subsection{Data Table}
 \label{s-data_table}

For convenience, we categorize the events according to their
microwave fluxes, and denote the groups similar to the GOES class:
mX (microwave-eXtreme) with $F_{35} > 10^4$ sfu, mS
(microwave-Strong) with $10^3$ sfu $ < F_{35} < 10^4$ sfu, mM
(microwave-Moderate) with $10^2$ sfu $< F_{35} < 10^3$ sfu, and mO
(microwave-Occulted).

The events and evaluated parameters of the bursts at 35 GHz along
with data on near-Earth protons are presented in
Table~\ref{tab:table1}. It lists chronologically the four groups
of events in the descending order of their microwave importance,
i.e., mX, mS, mM, and mO.

Column 1 of Table~\ref{tab:table1} presents the event number in
the table in the form No$^{\mathrm{Q}}$. The superscript, Q,
specifies the qualifier of the event defined as follows:

Q0 -- East event, no CME, no type II burst;

Q2 -- West event with a CME and a type II burst;

Q1 -- For all other events.

The qualifiers are only indicated for the events, which could be
observed by SOHO/LASCO, according to the SOHO LASCO CME catalog
(\cite{Yashiro2004},
http:/\negthinspace/cdaw.gsfc.nasa.gov/CME\_list).

Columns 2 and 3 indicate the date and time of the flare peak
according to GOES reports. Columns 4, 5, and 6 present the GOES
and H$\alpha$ importance of the flare and its coordinates.

Columns 7, 8, and 9 present the burst duration, the maximum flux
at 35 GHz in thousands of sfu (1 sfu = $10^{-22}$
W~m$^{-2}$~Hz$^{-1}$), and the microwave peak frequency.

Columns 10 and 11 list the maximum near-Earth total fluxes of
protons with $E > 100$~Mev ($J_{100}$) and with $ E > 10$~Mev
($J_{10}$). Column 12 characterizes the integral proton spectrum
with a parameter $\delta_{\mathrm{p}} =
\log_{10}(J_{10}/J_{100})$, which is calculated from the peak
fluxes of protons with different energies occurring at different
times, thus attempting to take account of their velocity
dispersion. Column 13 presents the magnitude of a GLE, if it has
occurred.

\begin{longtable}{rccllcrrrrrrr}
  \caption{Analyzed events}
  \label{tab:table1}
  \hline
\multicolumn{1}{c}{No$^\mathrm{Q}$}& \multicolumn{1}{c}{Date} &
\multicolumn{1}{c}{$T_{\mathrm{peak}}$}
 & \multicolumn{3}{c}{Flare}
 & \multicolumn{3}{c}{Microwave burst}
 &  & \multicolumn{3}{c}{Protons near Earth} \\

  &  &
  & \multicolumn{1}{c}{GOES} & \multicolumn{1}{c}{H$\alpha$} & \multicolumn{1}{c}{Position}
  & \multicolumn{1}{c}{$\Delta t_{\mathrm{35}}$} & \multicolumn{1}{c}{$F_{\max}$}
  & \multicolumn{1}{c}{$f_\mathrm{peak}$} & \multicolumn{1}{c}{$J_{100}$}
  & \multicolumn{1}{c}{$J_{10}$} & \multicolumn{1}{c}{$\delta_{\mathrm{p}}$}
  & \multicolumn{1}{c}{GLE} \\

  &  &  &  \multicolumn{3}{c}{}
  & \multicolumn{1}{c}{min} & \multicolumn{1}{c}{$10^3$ sfu}
  & \multicolumn{1}{c}{GHz} & \multicolumn{1}{c}{pfu} & \multicolumn{1}{c}{pfu}
  & \multicolumn{1}{c}{} & \multicolumn{1}{c}{\%} \\

  \hline
\multicolumn{1}{c}{1} & \multicolumn{1}{c}{2} &
\multicolumn{1}{c}{3} & \multicolumn{1}{c}{4} &
\multicolumn{1}{c}{5} & \multicolumn{1}{c}{6} &
\multicolumn{1}{c}{7} & \multicolumn{1}{c}{8} &
\multicolumn{1}{c}{9} & \multicolumn{1}{c}{10} &
\multicolumn{1}{c}{11} & \multicolumn{1}{c}{12} & \multicolumn{1}{c}{13} \\

\endfirsthead
  \hline
\endhead
  \hline
\endfoot
  \hline
\endlastfoot
  \hline

\multicolumn{13}{c}{mX events with extreme fluxes at 35 GHz ($F_{35}> 10^4$ sfu)} \\

 1 & 1990-04-15 & 02:59 & X1.4 & 2B & N32E54 & 66 & 20 & 11 & 0.04 & 9 & 2.4$^d$ & -- \\
 2 & 1990-05-21 & 22:15 & X5.5 & 2B & N34W37 & 7 & 38 & 47 & 18 & 300 & 1.22 & 24 \\
 3 & 1991-03-22 & 22:44 & X9.4 & 3B & S26E28 & 2 & 122 & 35 & 55 & 28000 & 2.70 & -- \\
 4 & 1991-03-29 & 06:45 & X2.4 & 3B & S28W60 & 7 & 11 & 30 & $<$0.1 & 20 & -- & -- \\
 5 & 1991-05-18 & 05:13 & X2.8 & 2N & N32W87 & 26 & 21 & 26 & $<$0.1 & 7 & -- & -- \\
 6 & 1991-06-04 & 03:41 & X12 & 3B & N30E60 & 15 & 130$^a$ & 44 & 2 & 50 & 1.40 & -- \\
 7 & 1991-06-06 & 01:09 & X12 & 3B & N33E44 & 17 & 130$^a$ & 46 & 2.5 & 200 & 1.90 & -- \\
 8 & 1991-06-09 & 01:39 & X10 & 3B & N34E04 & 7 & 74 & 36 & 1.2 & 80 & 1.82 & -- \\
 9 & 1991-06-11 & 02:06 & X12 & 3B & N32W15 & 18 & 46 & 30 & 42 & 2500 & 1.77 & 12 \\
10 & 1991-10-24 & 02:38 & X2.1 & 3B & S15E60 & 0.6 & 34 & 35 & -- & -- & -- & -- \\
11 & 1992-11-02 & 02:54 & X9 & 2B & S23W90 & 15 & 41 & 35 & 70 & 800 & 1.06 & 6.5 \\
12$^2$ & 2001-04-02 & 21:48 & X17 & --$^d$ & N18W82 & 6 & 25 & 35 & 4.8 & 380 & 1.90 & -- \\
13$^0$ & 2002-07-23 & 00:31 & X4.8 & 2B & S13E72 & 17 & 15 & 35 & -- & -- & -- & -- \\
14$^2$ & 2002-08-24 & 01:00 & X3.1 & 1F & S02W81 & 16 & 11 & 18 & 27 & 220 & 0.91 & 14 \\
15$^2$ & 2004-11-10 & 02:10 & X2.5 & 3B & N09W49 & 7 & $>$10$^{b,c}$ & $>$17$^{b,c}$ & 2 & 75 & 1.57 & -- \\
16$^2$ & 2005-01-20 & 06:46 & X7.1 & 2B & N12W58 & 25 & 85 & 28 & 680 & 1800 & 0.42 & 5400 \\
17$^2$ & 2006-12-13 & 02:40 & X3.4 & 4B & S06W24 & 31 & 14 & 45 & 88 & 695 & 0.89 & 92 \\
18$^1$ & 2012-03-07 & 00:24 & X5.4 & 3B & N17E15 & 80 & 11 & 17 & 67 & 1500 & 1.35 & -- \\
19$^2$ & 2012-07-06 & 23:08 & X1.1 & --$^d$ & S15W63 & 3 & 17 & 35 & 0.27 & 22 & 1.91 & -- \\

\multicolumn{13}{c}{mS events with strong fluxes at 35 GHz ($10^3 < F_{35} < 10^4$ sfu)} \\

20 & 1990-05-11 & 05:42 & X2.4 & SF & N15E13 & 14 & 2.0 & 15 & -- & -- & -- & --\\
21 & 1990-05-21 & 01:24 & M4.8 & 1B & N33W30 & 7 & 1.3 & 31 & -- & -- & -- & --\\
22 & 1990-05-23 & 04:20 & M8.7 & 1B & N33W55 & 10 & 1.0 & 10 & -- & -- & -- & --\\
23 & 1990-06-10 & 07:17 & M2.3 & 2B & N10W10 & 3 & 1.0 & 19 & -- & -- & -- & --\\
24 & 1991-01-25 & 06:32 & X10 & 1N & S12E90 & 6 & 9.4 & 28 & 0.3 & 1 & 0.52 & --\\
25 & 1991-03-05 & 23:26 & M6.2 & SF & S23E79 & 2 & 1.4 & 11 & -- & -- & -- & --\\
26 & 1991-03-07 & 07:49 & X5.5 & 3B & S20E62 & 3 & 2.0 & 32 & -- & -- & -- & --\\
27 & 1991-03-13 & 08:03 & X1.3 & 2B & S11E43 & 2 & 3.6 & 15 & 0.03 & 4.6 & 2.18 & --\\
28 & 1991-03-16 & 00:48 & X1.8 & 2B & S10E09 & 3 & 3.2 & 28 & -- & -- & -- & --\\
29 & 1991-03-16 & 21:52 & M6.0 & 2B & S09W04 & 4 & 1.6 & 21 & -- & -- & -- & --\\
30 & 1991-03-19 & 01:57 & M6.7 & 2B & S10W33 & 1 & 7.2 & 32 & -- & -- & -- & --\\
31 & 1991-03-21 & 23:39 & M5.4 & 2B & S25E40 & 3 & 7.2 & 32 & -- & -- & -- & --\\
32 & 1991-03-23 & 22:06 & M5.6 & 2B & S25E16 & 15 & 1.7 & 29 & -- & -- & -- & --\\
33 & 1991-03-25 & 00:17 & X1.1 & 2B & S26E01 & 11 & 3.9 & 18 & -- & -- & -- & --\\
34 & 1991-03-25 & 08:09 & X5.3 & 3B & S25W03 & 4 & 4.2 & 18 & 0.5 & 150 & 2.47 & --\\
35 & 1991-05-16 & 06:49 & M8.9 & 2B & N30W56 & 9 & 8.0 & 27 & -- & -- & -- & --\\
36 & 1991-05-29 & 23:43 & X1.0 & 2B & N05E38 & 1 & 1.7 & 21 & -- & 0.8 & -- & --\\
37 & 1991-06-30 & 02:56 & M5.0 & 1N & S06W19 & 0.8 & 2.0 & 20 & 0.2 & -- & -- & --\\
38 & 1991-07-30 & 07:07 & M7.2 & 1N & N14W58 & 0.9 & 2.0 & 30 & -- & -- & -- & --\\
39 & 1991-07-31 & 00:48 & X2.3 & 2B & S17E11 & 5 & 1.6 & 18 & -- & -- & -- & --\\
40 & 1991-08-02 & 03:13 & X1.5 & 2B & N25E15 & 8 & 1.2 & 13 & 0.15 & -- & -- & --\\
41 & 1991-08-03 & 01:22 & M2.9 & 1N & N24E05 & 3 & 2.8 & 25 & 0.15 & -- & -- & --\\
42 & 1991-08-25 & 00:51 & X2.1 & 2B & N24E77 & 29 & 1.4 & 10 & 0.03 & 21 & 2.84 & --\\
43 & 1991-10-27 & 05:42 & X6.1 & 3B & S13E15 & 6 & 8.8 & 12 & -- & 40 & -- & --\\
44 & 1991-11-02 & 06:45 & M9.1 & 2B & S13W61 & 3 & 1.4 & 11 & -- & 0.3 & -- & --\\
45 & 1991-11-15 & 22:37 & X1.5 & 3B & S13W19 & 4 & 1.5 & 20 & 0.28 & 1.1 & 0.59 & --\\

\hline

\multicolumn{7}{l}{$^a$Interpolated from data at 17 and 80 GHz} &
     \multicolumn{6}{l}{$^c$Estimated from lower-frequency data} \\
 \multicolumn{7}{l}{$^b$Estimated from NoRH data}  &
     \multicolumn{6}{l}{$^d$Uncertain} \\
 \multicolumn{7}{l}{$^0$East event, no CME, no type II burst} &
     \multicolumn{6}{l}{$^2$West event with CME and type II} \\
 \multicolumn{7}{l}{$^1$All other events} & \multicolumn{6}{l}{} \\

\end{longtable}

\setcounter{table}{0}

\begin{longtable}{rccllcrrrrrrr}
  \caption{(Continued.)}
  \hline
 \multicolumn{1}{c}{No$^\mathrm{Q}$} & \multicolumn{1}{c}{Date} & \multicolumn{1}{c}{$T_{\mathrm{peak}}$}
 & \multicolumn{3}{c}{Flare}
 & \multicolumn{3}{c}{Microwave burst}
 &  & \multicolumn{3}{c}{Protons near Earth} \\

  &  &
  & \multicolumn{1}{c}{GOES} & \multicolumn{1}{c}{H$\alpha$} & \multicolumn{1}{c}{Position}
  & \multicolumn{1}{c}{$\Delta t_{\mathrm{35}}$} & \multicolumn{1}{c}{$F_{\max}$}
  & \multicolumn{1}{c}{$f_\mathrm{peak}$} & \multicolumn{1}{c}{$J_{100}$}
  & \multicolumn{1}{c}{$J_{10}$} & \multicolumn{1}{c}{$\delta_{\mathrm{p}}$}
  & \multicolumn{1}{c}{GLE} \\

  &  &  &  \multicolumn{3}{c}{}
  & \multicolumn{1}{c}{min} & \multicolumn{1}{c}{$10^3$ sfu}
  & \multicolumn{1}{c}{GHz} & \multicolumn{1}{c}{pfu} & \multicolumn{1}{c}{pfu}
  & \multicolumn{1}{c}{} & \multicolumn{1}{c}{\%} \\

  \hline
\multicolumn{1}{c}{1} & \multicolumn{1}{c}{2} &
\multicolumn{1}{c}{3} & \multicolumn{1}{c}{4} &
\multicolumn{1}{c}{5} & \multicolumn{1}{c}{6} &
\multicolumn{1}{c}{7} & \multicolumn{1}{c}{8} &
\multicolumn{1}{c}{9} & \multicolumn{1}{c}{10} &
\multicolumn{1}{c}{11} & \multicolumn{1}{c}{12} & \multicolumn{1}{c}{13} \\

\endfirsthead
  \hline
\endhead
  \hline
\endfoot
  \hline
\endlastfoot
  \hline
46 & 1992-02-14 & 23:07 & M7.0 & 2B & S12E02 & 1 & 1.0 & 12 & -- & -- & -- & --\\
47 & 1992-02-27 & 08:08 & C2.6 & SF & N03W05 & 0.6 & 1.2 & 20 & -- & -- & -- & --\\
48 & 1992-06-28 & 04:54 & X1.8 & SF & N11W90 & 14 & 1.3 & 10 & 0.22 & 14 & 1.8 & --\\
49 & 1994-01-16 & 23:17 & M6.1 & 1N & N07E71 & 9 & 1.2 & 35 & -- & -- & -- & --\\
50$^2$& 1997-11-04 & 05:57 & X2.1 & 2B & S14W33 & 3 & 1.0 & 18 & 2.3 & 72 & 1.5 & --\\
51 & 1998-08-08 & 03:15 & M3.0 & --$^d$ & N14E72 & 0.7 & 2.0 & 24 & -- & -- & -- & --\\
52 & 1998-08-22 & 00:01 & M9.0 & 2B & N42E51 & 6 & 1.0 & 18 & -- & 2.5 & -- & --\\
53$^2$ & 1998-11-22 & 06:38 & X3.7 & 1N & S27W82 & 7 & 6.7 & 20 & 0.22 & 4 & -- & --\\
54$^1$ & 1999-08-20 & 23:06 & M9.8 & 1N & S23E60 & 1 & 3.0 & 29 & -- & -- & -- & --\\
55$^2$ & 1999-12-28 & 00:43 & M4.5 & 2B & N23W47 & 2 & 2.2 & 14 & 0.1 & 0.5 & 0.69 & --\\
56$^1$ & 2000-09-30 & 23:19 & X1.2 & SF & N09W75 & 4 & 5.2 & 29 & -- & -- & -- & --\\
57$^2$ & 2000-11-24 & 04:59 & X2.0 & 3B & N19W05 & 2 & 9.3 & 32 & 0.58 & 8 & 1.13 & --\\
58$^2$ & 2001-03-10 & 04:03 & M6.7 & 1B & N26W42 & 1 & 1.6 & 24 & -- & 0.2 & -- & --\\
59$^1$ & 2001-04-03 & 03:36 & X1.2 & 1N & S21E71 & 31 & 2.9 & 11 & 0.1 & 100 & 3 & --\\
60$^2$ & 2001-04-10 & 05:32 & X2.3 & 3B & S24W05 & 30 & 2.9 & 9 & 0.47 & 355 & 2.87 & --\\
61$^0$ & 2001-10-12 & 03:23 & C7.6 & SF & N16E70 & 1 & 1.3 & 40 & -- & -- & -- & --\\
62$^1$ & 2001-10-25 & 05:18 & C5.2 & SF & S19W17 & 1 & 1.2 & 25 & -- & 1 & -- & --\\
63$^2$ & 2002-02-20 & 06:11 & M5.1 & 1N & N13W68 & 5 & 1.5 & 27 & 0.1 & -- & -- & --\\
64$^1$ & 2002-07-18 & 03:32 & M2.2 & SB & N19W27 & 2 & 1.4 & 20 & -- & -- & -- & --\\
65$^1$ & 2002-08-20 & 01:49 & M5.0 & SF & S08W34 & 0.5 & 1.8 & 40 & 0.07 & -- & -- & --\\
66$^2$ & 2002-08-21 & 01:38 & M1.4 & SF & S10W47 & 1 & 1.3 & 26 & -- & -- & -- & --\\
67$^2$ & 2002-08-21 & 05:31 & X1.0 & 1B & S09W50 & 0.7 & 1.4 & 27 & -- & -- & -- & --\\
68$^1$ & 2003-04-26 & 00:55 & M2.1 & SF  & N20W65 & 2 & 2.2 & 30 & -- & -- & -- & --\\
69$^1$ & 2003-04-26 & 03:03 & M2.1 & SN & N20W69 & 0.3 & 2.4 & 30 & -- & -- & -- & --\\
70$^2$ & 2003-05-28 & 00:26 & X3.6 & 1B & S08W22 & 14 & 3.4 & 16 & 0.15 & 121 & 2.9 & --\\
71$^2$ & 2003-05-29 & 01:01 & X1.2 & 2B & S07W31 & 12 & 1.2 & 14 & -- & -- & -- & --\\
72$^2$ & 2003-05-31 & 02:21 & M9.3 & 2B & S06W60 & 8 & 1.7 & 15 & 0.8 & 27 & 1.53 & \\
73$^1$ & 2003-06-15 & 23:44 & X1.3 & SF & S07E80 & 8 & 1.9 & 11 & -- & 0.3 & -- & --\\
74$^1$ & 2003-06-17 & 22:53 & M6.8 & --$^d$ & S08E58 & 23 & 1.8 & 30 & 0.02 & 16 & 2.9 & --\\
75$^1$ & 2003-10-24 & 02:46 & M7.6 & 1N & S19E72 & 32 & 3.9 & 30 & -- & -- & -- & --\\
76$^1$ & 2003-10-26 & 06:14 & X1.2 & 3B & S17E42 & 62 & 3.6 & 17 & -- & -- & -- & --\\
77 & 2004-01-06 & 06:22 & M5.8 & --$^d$ & N05E89 & 8 & 1.0 & 12 & -- & -- & -- & --\\
78$^1$ & 2004-01-07 & 03:59 & M4.5 & 2N & N02E82 & 9 & 1.8 & 40 & -- & -- & -- & --\\
79$^0$ & 2004-07-16 & 02:03 & X1.3 & --$^d$ & S10E39 & 5 & 1.5 & 20 & -- & -- & -- & --\\
80$^1$ & 2004-08-14 & 05:43 & M7.4 & 2N & S12W29 & 7 & 1.1 & 20 & -- & -- & -- & --\\
81$^2$ & 2004-10-30 & 06:14 & M4.2 & SF & N13W21 & 7 & 1.3 & 20 & 0.04 & 0.9 & 1.35 & --\\
82$^1$ & 2004-11-03 & 03:30 & M1.6 & 1N & N07E46 & 10 & 1.1 & 9 & -- & 0.4 & -- & --\\
83$^1$ & 2005-01-01 & 00:29 & X1.7 & --$^d$ & N04E35 & 6 & 1.7 & 15 & -- & -- & -- & --\\
84$^0$ & 2005-01-15 & 00:41 & X1.2 & 1B & N13E05 & 6 & 3.3 & 20 & -- & -- & -- & --\\
85$^1$ & 2005-07-30 & 06:25 & X1.3 & 2B & N11E58 & 27 & 1.1 & 10 & -- & -- & -- & --\\
86$^1$ & 2005-08-25 & 04:38 & M6.4 & 1N & N08E82 & 5 & 4.3 & 26 & -- & -- & -- & --\\
87$^0$ & 2005-09-13 & 23:21 & X1.7 & 1B & S11E10 & 6 & 5.0 & 36 & -- & 90 & -- & --\\
88$^1$ & 2005-09-17 & 06:05 & M9.8 & 2N & S11W41 & 6 & 1.3 & 25 & -- & 1.4 & -- & --\\
89$^2$ & 2011-08-04 & 03:57 & M9.3 & 2B & N16W49 & 11 & 1.4 & 11 & 1.5 & 77 & 1.71 & --\\
90$^2$ & 2011-08-09 & 08:05 & X6.9 & 2B & N17W83 & 6 & 1.0 & 14 & 2.5 & 22 & 0.94 & --\\
91$^2$ & 2012-01-23 & 03:59 & M8.7 & 2B & N29W36 & 39 & 2.0 & 4 & 2.3 & 2700 & 3.07 & --\\
\hline

\multicolumn{7}{l}{$^a$Interpolated from data at 17 and 80 GHz} &
     \multicolumn{6}{l}{$^c$Estimated from lower-frequency data} \\
 \multicolumn{7}{l}{$^b$Estimated from NoRH data}  &
     \multicolumn{6}{l}{$^d$Uncertain} \\
 \multicolumn{7}{l}{$^0$East event, no CME, no type II burst} &
     \multicolumn{6}{l}{$^2$West event with CME and type II} \\
 \multicolumn{7}{l}{$^1$All other events} & \multicolumn{6}{l}{} \\

\end{longtable}

\setcounter{table}{0}

\begin{longtable}{rccllcrrrrrrr}
  \caption{(Continued.)}
  \hline
\multicolumn{1}{c}{No$^\mathrm{Q}$}&\multicolumn{1}{c}{Date}&\multicolumn{1}{c}{$T_{\mathrm{peak}}$}
 & \multicolumn{3}{c}{Flare}
 & \multicolumn{3}{c}{Microwave burst}
 &  & \multicolumn{3}{c}{Protons near Earth} \\

  &  &
  & \multicolumn{1}{c}{GOES} & \multicolumn{1}{c}{H$\alpha$} & \multicolumn{1}{c}{Position}
  & \multicolumn{1}{c}{$\Delta t_{\mathrm{35}}$} & \multicolumn{1}{c}{$F_{\max}$}
  & \multicolumn{1}{c}{$f_\mathrm{peak}$} & \multicolumn{1}{c}{$J_{100}$}
  & \multicolumn{1}{c}{$J_{10}$} & \multicolumn{1}{c}{$\delta_{\mathrm{p}}$}
  & \multicolumn{1}{c}{GLE} \\

  &  &  &  \multicolumn{3}{c}{}
  & \multicolumn{1}{c}{min} & \multicolumn{1}{c}{$10^3$ sfu}
  & \multicolumn{1}{c}{GHz} & \multicolumn{1}{c}{pfu} & \multicolumn{1}{c}{pfu}
  & \multicolumn{1}{c}{} & \multicolumn{1}{c}{\%} \\

  \hline
\multicolumn{1}{c}{1} & \multicolumn{1}{c}{2} &
\multicolumn{1}{c}{3} & \multicolumn{1}{c}{4} &
\multicolumn{1}{c}{5} & \multicolumn{1}{c}{6} &
\multicolumn{1}{c}{7} & \multicolumn{1}{c}{8} &
\multicolumn{1}{c}{9} & \multicolumn{1}{c}{10} &
\multicolumn{1}{c}{11} & \multicolumn{1}{c}{12} & \multicolumn{1}{c}{13} \\

\endfirsthead
  \hline
\endhead
  \hline
\endfoot
  \hline
\endlastfoot
  \hline

\multicolumn{13}{c}{mM events with strong proton fluxes ($J_{100} > 10$ pfu, $10^2 < F_{35} < 10^3$ sfu)} \\

92$^2$ & 2000-11-08 & 23:28 & M7.8 & 1N & N10W75 & 53 & 0.14 & 2.8 & 320  & 14000 & 1.64 & --\\
93$^2$ & 2001-12-26 & 05:40 & M7.1 & 1B & N08W54 & 26 & 0.78 & 6.9 & 47 & 700 & 1.17 & 13\\
94$^2$ & 2002-04-21 & 01:15 & X1.5 & 1F & S14W84 & 83 & $\sim$0.4$^{a,b}$ & 5 & 20 & 2000 & 2.00 & --\\
95$^2$ & 2012-05-17 & 01:47 & M5.1 & 1F & N09W74 & 17 & 0.2 & 10 & 18 & 230 & 1.11 & 16\\

\multicolumn{13}{c}{mO backside events with strong proton fluxes ($J_{100} > 10$ pfu)} \\

96 & 1990-05-28 & 04:33$^d$ & -- & -- & N36W120 & 8 & 0.1 & 1.4 & 43 & 430 & 1.00 & 6 \\
97 & 2001-04-18 & 02:15$^d$ & C2.2 & -- & S20W115 & 4 & -- & -- & 12 & 230 & 1.28 & 26 \\
98 & 2001-08-15 & --$^d$ & -- & -- & W$>120$ & -- & -- & -- & 27 & 470 & 1.24 & -- \\

\hline

 \multicolumn{7}{l}{$^a$Interpolated from data at 17 and 80 GHz} &
     \multicolumn{6}{l}{$^c$Estimated from lower-frequency data} \\
 \multicolumn{7}{l}{$^b$Estimated from NoRH data}  &
     \multicolumn{6}{l}{$^d$Uncertain} \\
 \multicolumn{7}{l}{$^0$East event, no CME, no type II burst} &
     \multicolumn{6}{l}{$^2$West event with CME and type II} \\
 \multicolumn{7}{l}{$^1$All other events} & \multicolumn{6}{l}{} \\

\end{longtable}

\section{Results}
 \label{s-results}

\subsection{General Outcome}
 \label{s-outcome}

Table~\ref{tab:table1} demonstrates that most mX events produced
SEP events indeed: 89\% of both west and east mX events (total 19)
produced proton enhancements, with $J_{100} > 1$ pfu for 68\% of
the 19 events. Considerable proton fluxes were observed after the
events with very strong bursts at 35 GHz, even with their rather
far east location. A reduced proton productivity had two west
events No. 4 and 5, after which enhancements were distinct for
$>10$~MeV protons only. About 30\% of the mX events produced GLEs.
In these events, the proton spectrum indices were
$\delta_{\mathrm{p}} < 2$, and the microwave peak frequencies
$f_{\mathrm{peak}} \geq 18$ GHz (32 GHz on average over the 19
events).

The proton productivity of mS events is considerably lower. None
of them produced a GLE. Totally 52\% of the mS events (both west
and east) were not followed by any detectable enhancement even of
$> 10$ MeV protons.

\begin{figure} 
  \begin{center}
    \FigureFile(85mm,60mm){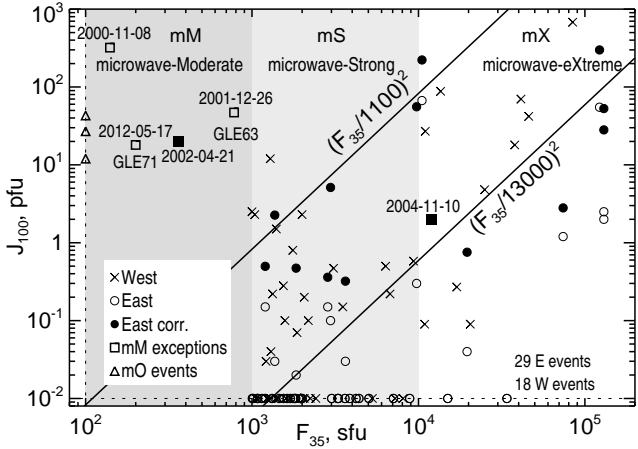}
  \end{center}
  \caption{Fluxes of $> 100$ MeV protons vs. radio fluxes at 35 GHz.
The filled squares denote the events with increased uncertainties
of 35~GHz fluxes. The black solid lines are arbitrarily chosen to
verify a direct relation between the observed $F_{35}$ and
$J_{100}$.}
 \label{fig:event_distribution}
 \end{figure}

To make the results clearer, Figure~\ref{fig:event_distribution}
presents the data from Table~\ref{tab:table1} as peak proton
fluxes with $E > 100$~MeV vs. peak microwave fluxes at 35 GHz. The
events without detectable proton fluxes falling outside the region
of the plot are presented at the horizontal dotted line
corresponding to $10^{-2}$ pfu to show their amount. Total of 18
west events and 29 east events have not produced detectable SEP
fluxes. Similarly, three backside events with microwave fluxes
$F_{35} < 100$~sfu (triangles) are shown at the vertical dotted
line corresponding to 100~sfu.

The SEP fluxes, whose sources have eastern locations (empty
circles), are known to be reduced due to the bend of the Parker
spiral deflecting charged particles away from the Earth. We have
additionally shown these events with the filled circles by
applying an empirical correction for the dependence of
$\exp^{-[(\lambda-54)/63]^2}$ on the longitude $\lambda$ (A.~Belov
2012, private communication) to partially compensate for the
reduction. The plot in Figure~\ref{fig:event_distribution}
suggests that this correction might be probably somewhat
overestimated.

The following groups of events are distinct in
Figure~\ref{fig:event_distribution}.

\begin{enumerate}

 \item
Events exhibiting a direct tendency between $F_{35}$ and
$J_{100}$, which occupy a wide band from the lower left corner to
the upper right one (`main sequence').

 \item
Strong bursts at 35 GHz without SEPs, schematically shown along
the lower horizontal dotted line.

 \item
Three big SEP events associated with backside sources (mO),
schematically shown along the left vertical dotted line.

 \item
Four exceptional mM events with large protons fluxes (squares in
the upper left region).

\end{enumerate}

\textit{Group 1}. Events of the `main sequence' show a general
correspondence between the proton fluxes and the $F_{35}$ fluxes,
being mostly within the band bounded with rather arbitrary lines
of $(F_{35}/1100)^2$ and $(F_{35}/13000)^2$, which reflect a
direct flare--SEP relation. The scatter is large for obvious
reasons. For example, flare-accelerated protons are affected by
escape conditions from active regions; shock-accelerated protons
are influenced by the plenitude of a seed population; and all
depend on the Sun--Earth connections.

\begin{figure}  
  \begin{center}
    \FigureFile(85mm,60mm){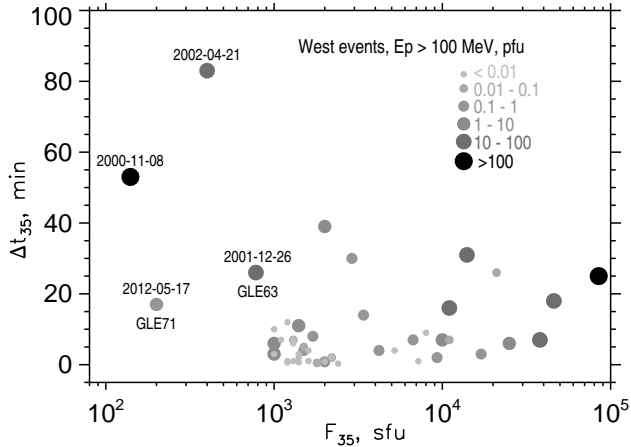}
  \end{center}
  \caption{Distribution of $>100$~Mev SEP events associated with west
source regions vs. peak fluxes and durations of 35 GHz bursts. The
peak proton fluxes are coded by the size and color of the filled
circles. The dates of the four mM exceptions are specified.}
 \label{fig:durations}
 \end{figure}

\textit{Group 2}. Two well-known reasons can account for the
absence of SEPs in these events. Those are east location, which
was already mentioned, and short duration. Poor proton production
in short-duration `impulsive' events established a long time ago
is usually interpreted by different acceleration mechanisms in
contrast to `gradual' events (see, e.g.,
\cite{Croom71,Cliver1989};
\authorcite{Reames2009} \yearcite{Reames2009,Reames2013}; and
references therein). Possible additional reasons for the
differences between these two categories will be discussed in
section~\ref{s-discussion}. To make the situation with the
durations in the considered set clearer, we have plotted west
events on the $(F_{35} - \Delta t_{35})$ plane in
Figure~\ref{fig:durations} with coding their SEP importance by
different size and color of the filled circles. Indeed, low SEP
fluxes were observed in short-duration events concentrated at the
bottom part of the plot. We remind that the burst durations at 35
GHz are generally shorter than for longer-wave bursts and always
shorter than in soft X-rays, which are usually considered.

\textit{Group 3}. No conclusion can be drawn for the three mO
backside events for the lack of information about microwave
sources occulted by the limb. The soft X-ray and microwave fluxes
ascribed to some of these events might be inadequate.

\textit{Group 4}. The four mM exceptions with moderate microwave
flux and incomparably strong SEP events are located high above the
`main sequence'. Two of these events produced GLEs: 2001 December
26 (GLE63) and the recent event of 2012 May 17 (GLE71). Two
remaining events, 2000 November 8 and 2002 April 21, are also well
known. The SEP spectra in these events $\delta_{p} \leq 2$ were
relatively hard, with $\delta_{p} < 1.2$ in two GLE events, being
atypical of non-flare-related filament eruptions, where
$\delta_{p}\sim 3 $ \citep{Chertok09}. There are no obvious
indications of reasons for the exceptional characteristics of
these events. Significance of group~4 is supported by the absence
of $J_{100}> 10$~pfu enhancements among the mS events.

\subsection{Relation to SEP Spectra}

The SEP spectra were relatively hard, $\delta_{p} \leq 2$, in most
west events with high $f_\mathrm{peak}
> 20$ GHz. There is a weak tendency of hardening the proton spectra
(decrease of $\delta_{\mathrm{p}}$) with increase of the microwave
peak frequency $f_{\mathrm{peak}}$. This tendency is consistent
with the conclusions of \citet{Chertok09}. Statistics of SEP
events with known $\delta_{\mathrm{p}}$ in Table~\ref{tab:table1}
is relatively poor to figure out this tendency with confidence. We
therefore involve additional information from \citet{Chertok09}
about SEP events and radio bursts recorded with the USAF RSTN
network during 1988--2006. These data are available at
http:/\negthinspace
/www.izmiran.ru/$\sim$ichertok/SEPs\_radio/Table.html.

To minimize the influence of the heliolongitude on the SEP
spectrum, the events associated with west solar sources only are
analyzed. We compare two subsets of SEP events: i)~SEP events with
corresponding NoRP mX and mS bursts ($F_{35} > 10^3$ sfu) from
Table~\ref{tab:table1}, and ii)~events without strong bursts at 35
GHz selected from \citet{Chertok09}. Subset (ii) contains only
events, in which $F_{35} < 10^3$ sfu judging from lower-frequency
RSTN data.

\begin{figure}  
  \begin{center}
    \FigureFile(85mm,68mm){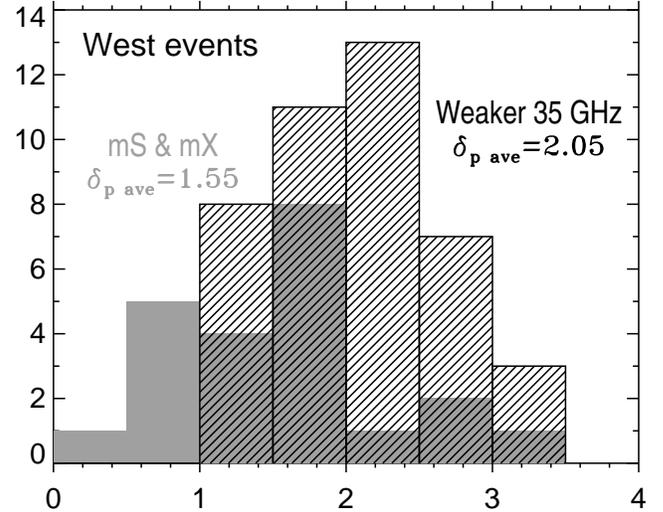}
  \end{center}
  \caption{Histograms of the protons energy index $\delta_{p}$ for
SEP events with strong NoRP 35 GHz bursts (mS and mX, shaded) and
for SEP events with weaker RSTN bursts (line-filled).}
 \label{fig:histogram}
 \end{figure}

The distribution of $\delta_{\mathrm{p}}$ in the two subsets is
shown in Figure~\ref{fig:histogram}. The shaded histogram presents
SEP events with certain $\delta_{p}$ corresponding to NoRP mX and
mS bursts from west solar sources. These are events No. 2, 9, 11,
12, 14, 15, 16, 17, 19, 34, 45, 48, 50, 55, 57, 60, 70, 72, 81,
89, 90, and 91 (totally 22 events). The line-filled histogram
presents SEP events (also from west solar sources), which were not
preceded by strong bursts at 35 GHz (totally 42 events, including
the four exceptional mM events No. 92--95).

Indeed, the SEP events after strong 35 GHz bursts had
predominantly harder spectra with an average
$\overline{\delta_{p}} = 1.55$. The histogram for the second
subset with an average $\overline{\delta_{p}} = 2.05$ is
apparently shifted right. Exclusion of the four exceptional mM
events (No. 92--95, $\delta_{p} = 1.11-2.00$, squares in the upper
left region of Figure~\ref{fig:event_distribution}) from the
histogram increases $\overline{\delta_{p}}$ to 2.11 in the second
subset.

\section{Discussion}
 \label{s-discussion}

The characteristics of the mX events, whose microwave peak
frequencies reach very high values (32 GHz on average), suggest
flaring above the sunspot umbrae, where strongest magnetic fields
are reached. This conclusion follows from properties of
gyrosynchrotron emission
\citep{DulkMarsh1982,Stahli1989,Krucker2013}, as the next section
will confirm. The strong dependence of the energy release rate on
the magnetic field strength is expected from the standard flare
model, as \authorcite{Asai2002} (\yearcite{Asai2002,Asai2004})
have shown. \citet{Grechnev2008} and \citet{Kundu2009} also
demonstrated extreme parameters of sunspot-associated flares such
as strong hard X-ray and gamma-ray emissions and high SEP
productivity.

\begin{figure*}  
  \begin{center}
    \FigureFile(170mm,109mm){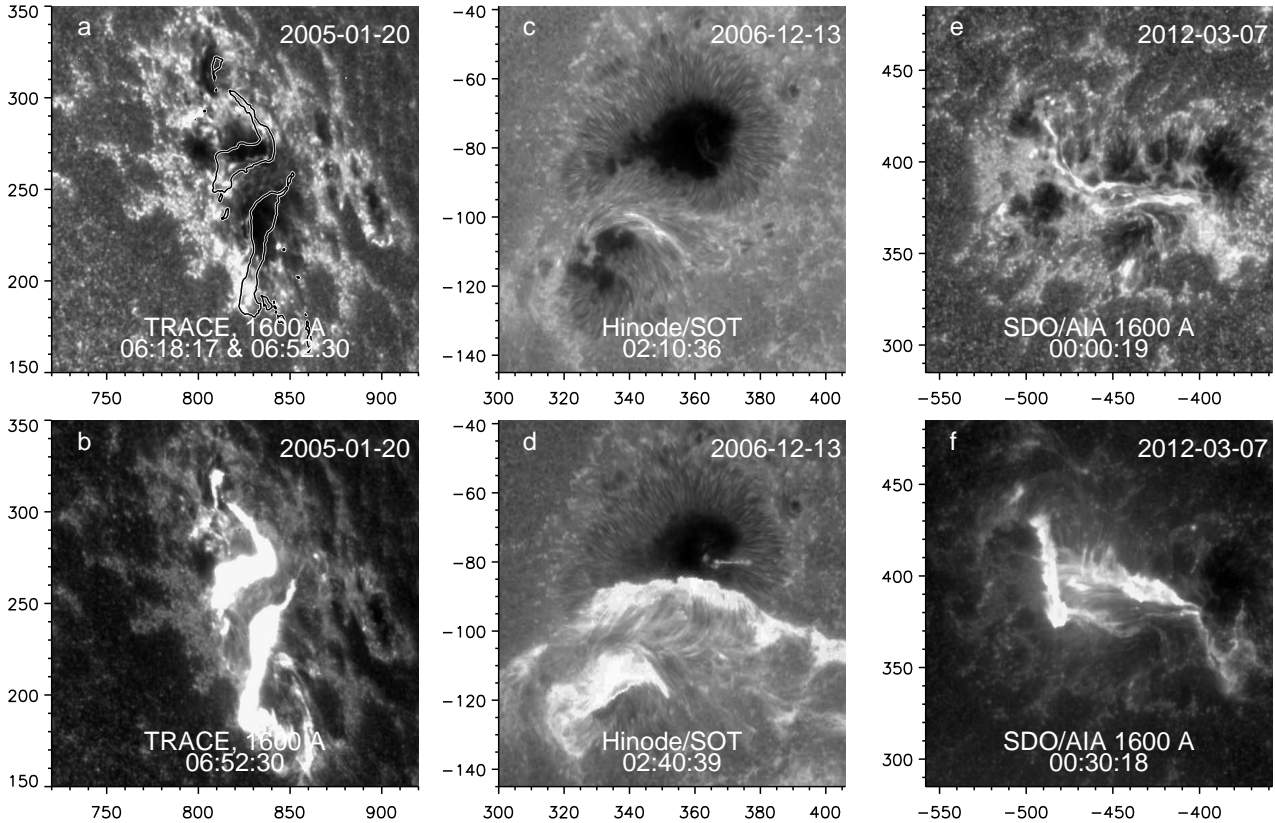}
  \end{center}
  \caption{Three sunspot-associated flares on 2005 January 20 (a, b),
on 2006 December 13 (c, d), and on 2012 March 07 (e, f). Top:
before the flare, bottom: during the flare. The contours overlayed
on the image in panel (a) outline the ribbons visible in panel
(b). The axes show hereafter arc seconds from the solar disk
center.}
 \label{fig:umbrae_ribbons}
 \end{figure*}

Figure~\ref{fig:umbrae_ribbons} confirms this assumption for three
extreme events: 2005 January 20 (a, b; X7.1, No.~16, GLE69), 2006
December 13 (c, d; X3.4, No.~17, GLE70), and 2012 March 07 (e, f;
X5.4, No.~18). The ribbons crossed the sunspot umbrae also
in a series of big white-light flares, which occurred on 1991 June
4, 9, and 11 (\cite{Sakurai92}; events No. 6, 8, and 9 in
Table~\ref{tab:table1}). Flaring above the sunspot umbrae seems to
be typical of mX and some mS events indeed.

\subsection{Well-Sampled Event of 2001 August 25}

To provide further verification of the assumption about the
relation of strong high-frequency bursts with flaring in strong
magnetic fields above the sunspot umbrae, we briefly consider an
extreme white-light flare of 2001 August 25 (X5.3/3B, S17~E34).
This flare \citep{Metcalf2003} was responsible for a big neutron
event, extreme hard X-ray and gamma-ray emissions
\citep{Watanabe2003,Kuznetsov2006,Livshits2006}, and a fast CME.
The SEP event was not pronounced at Earth due to the east location
of the solar region. The event has occurred during the nighttime
in Nobeyama; nevertheless, we use NoRH data to estimate a probable
magnetic field strength in the flaring region.

This flare was chosen for a unique coverage of its spectrum at 1--18
GHz (Owens Valley Solar Array, OVSA), at 89.4 GHz (the nulling
interferometer at Bern University), and at 212 and 405 GHz (Solar
Submillimeter-wave Telescope, SST). The radio measurements from
microwaves up to submillimeters were addressed by
\citet{Raulin2004}, who concluded that the gyrosynchrotron emission
up to $\sim 10^5$ sfu was produced by electrons radiating in a
1000--1100~G region, because `\textit{magnetic fields higher than
1100 G should be excluded since they produce a peak frequency at or
above 90 GHz, which was not observed}'. The authors mentioned a
possibility of an inhomogeneous source, but their model did not
include it. \citet{Krucker2013} confirm the gyrosynchrotron
mechanism of the submillimeter emission in this event.

\begin{figure}
  \begin{center}
    \FigureFile(85mm,68mm){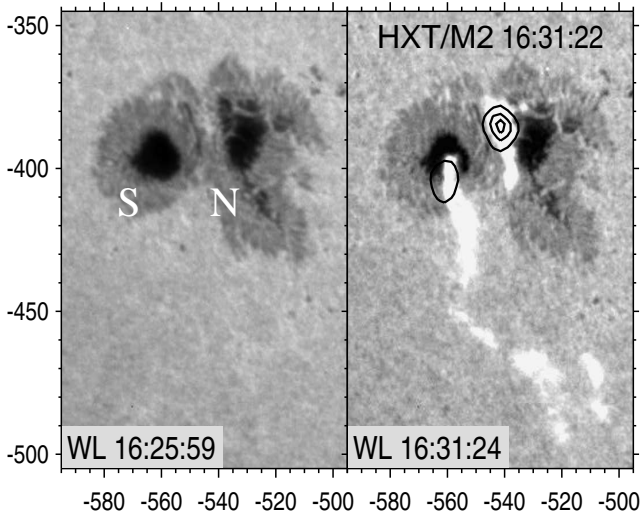}
  \end{center}
  \caption{The flare configuration in the 2001 August 25 extreme event.
The gray scale background presents white light images (TRACE)
before the flare (left) and during the flare peak (right).
Contours in the right panels present HXR sources (Yohkoh/HXT/M2,
33--53 keV). The contour levels are [0.2, 0.5, 0.8] of the maximum
value. The magnetic polarities of the sunspots are indicated in
the left panel.}
 \label{fig:20010825_sources}
 \end{figure}

Figure~\ref{fig:20010825_sources} shows TRACE white-light images
of the active region before the flare (left) and the flare
configuration near the peak of the event (right). The east flare
ribbon intruded far into the umbrae of the east S-polarity
sunspot. The west ribbon covered the edge of the west N-polarity
sunspot's umbra. The black contours present hard X-ray sources,
whose images we have produced from Yohkoh/HXT data in the 33--53
keV M2 channel \citep{Kosugi1991}.

The flare configuration with the ribbons above the umbrae
corresponds to our expectations. Although \citet{Raulin2004}
inferred strong magnetic fields in the gyrosynchrotron source, we
will reconsider the conclusion of the authors about the field
strength in this extreme event by using a simple model of an
inhomogeneous gyrosynchrotron source described by
\citet{Kundu2009}.

\begin{figure}
  \begin{center}
    \FigureFile(85mm,85mm){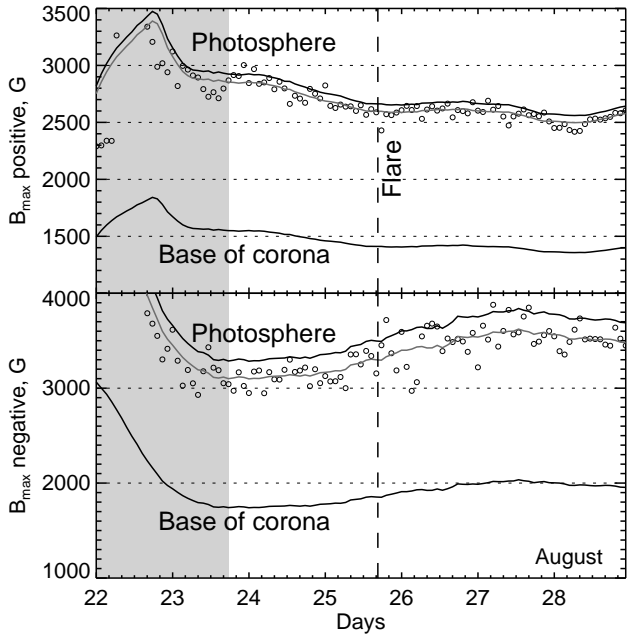}
  \end{center}
  \caption{Estimation of the probable magnetic field strength
from one-week-long MDI magnetograms and NoRH observations. The
circles present the maximum fields strength in each sunspot
measured from radialized 96-min magnetograms (top N-polarity,
bottom S-polarity). The gray curves show their smoothing. The
upper black envelopes of the measured points present evolution of
the probable maximum field at the photosphere. The magnified lower
black curves correspond to the base of the corona. The
radialization in the shaded interval is most likely unrealistic.}
 \label{fig:mag_strength}
 \end{figure}

The first point in question is a probable magnetic field strength
in the corona, where the flare source was located. To estimate it,
we have analyzed the maximum field strengths in both sunspots from
a one-week-long set of 96-min MDI magnetograms and referred them
to NoRH observations. The secant correction was applied to the
magnetograms (the `zradialize' SolarSoft routine). The
magnetograms suffer from the `high-field saturation', especially
in the east sunspot of the S-polarity.
Figure~\ref{fig:mag_strength} presents the maximum field strengths
measured in the positive west sunspot (top) and the negative east
sunspot (bottom) with small circles. The `saturation' is
especially pronounced in the larger scatter of the bottom plot.

The gray curves present a boxcar smoothing of the measured points
over 15 neighbors. The plots show an unrealistic rise of the
magnetic field strength near the limb before 23--24 August
suggestive of an excessive secant correction. We consider the shaded
region to be spurious.

The upper black curves approximately enveloping the measured
points were computed by magnifying the gray curves by factors of
1.025 for the N-polarity sunspot and 1.060 for the S-polarity
sunspot. They present the evolution of a probable maximum field at
the photospheric level, which might be still underestimated.
However, the magnetic field strength in the corona is important.
To estimate it, we note that NoRH observations at 17 GHz on August
27--29 reveal a sunspot-associated source above the east sunspot
with a brightness temperature of 0.2--0.5 MK and the degree of
polarization $> 50\%$. These properties are typical of the
gyroresonance emission, which occurs in sunspot-associated sources
at 17 GHz at the third harmonic of the gyrofrequency, i.e., in
magnetic fields of $\approx 2000$~G \citep{Vourlidas2006}. The
lower curves in Figure~\ref{fig:mag_strength} are referred to this
estimate and calculated as the upper envelopes scaled by a factor
of 0.53. We get probable magnetic field strengths at the base of
the corona during the flare (the vertical dashed line) of about
$+1400$~G in the west sunspot and $-1800$~G in the east sunspot.

\begin{figure}
  \begin{center}
    \FigureFile(85mm,85mm){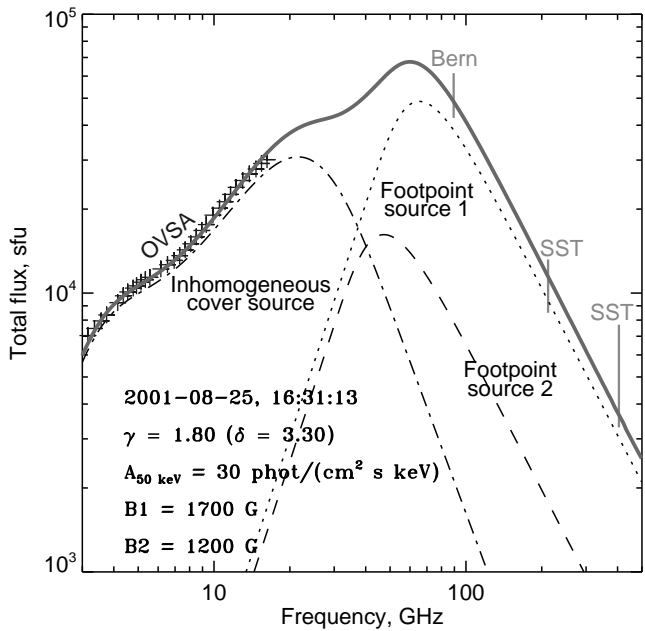}
  \end{center}
  \caption{The gyrosynchrotron spectrum of the 2001 August 25 extreme
flare. The symbols present the observations of OVSA (crosses),
Bern, and SST (gray bars). The thick curve is calculated from the
model. The dotted and dashed curves show the spectra of two
footpoint kernel sources. The dash-dotted curve presents the
inhomogeneous cover source in the top part of the magnetic
configuration.}
 \label{fig:modeled_spectrum}
 \end{figure}

Figure~\ref{fig:modeled_spectrum} presents with the symbols the
spectrum observed close to the flare peak (from \cite{Raulin2004})
along with the model one presented with the thick line. The
three-component model \citep{Kundu2009} simulates emission from
two footpoint sources (dotted and dashed) visible through an
inhomogeneous frequency-dependent cover source (dash-dotted; see
\cite{Bastian98}) based on the expressions from
\citet{DulkMarsh1982} and \citet{White2011}. The parameters of the
HXR spectrum were evaluated by V.~Kurt from data of CORONAS-F/SONG
\citep{Kuznetsov2011} and Yohkoh/GRS \& HXT
\citep{Yoshimori1991,Kosugi1991}: $\gamma = 2.0$, $A_{50 \
\mathrm{keV}} = 30$ photons~cm$^{-2}$~s$^{-1}$~keV$^{-1}$. A small
difference between the index $\gamma = 1.8$ used in the model and
the actual $\gamma = 2.0$ can be due to the spectral hardening of
trapped electrons \citep{MelnikovMagun98,Silva2000,Kundu2009}. The
size of each source was estimated from the HXT/M2 images in
Figure~\ref{fig:20010825_sources}. The magnetic field strengths in
the sources were taken $-1700$~G and $+1200$~G.

The model correctly reproduces main features of the observed
spectrum. Invoking the inhomogeneous source removes the limitation
of B $\leq 1100$ G, which restrained considerations of
\citet{Raulin2004}: the magnetic field in the source region could
be considerably stronger than the authors concluded. The very
strong magnetic field in the flare region determined the extreme
properties of this big flare, indeed. This example also shows that
the turnover frequency of the gyrosynchrotron spectrum can be
displaced left due to the contribution of a large inhomogeneous
cover source, thus affecting column~9 in Table~\ref{tab:table1}.
The displacement can be large and hide an indication of strong
fields in the lowest part of the flare configuration
\citep{White2003,Kundu2009}.

\subsection{Exceptional mM Events}
The four exceptional mM events from group~4 (squares in the upper
left part of Figure~\ref{fig:event_distribution}) look
challenging: large proton enhancements $J_{100} > 10$~pfu occurred
in association with moderate bursts, in which $F_{35} < 1000$~sfu.
The peak frequencies of the microwave spectra in these events were
below 10 GHz. The `Big Flare Syndrome' and the presumable
exclusive responsibility of the shock-acceleration do not clarify
the situation: while a fast CME and strong shock are expected
after a big flare, such expectations from a moderate flare seem
unlikely (see section~\ref{s-introduction}).

The relation between the microwave fluxes from the four
exceptional events and their proton productivity appears to be
distorted for some reasons. The first possibility is prompted by
the locations of the mM exceptions in
Figure~\ref{fig:event_distribution} close to the occulted mO
events, i.e., possible contributions from nearly simultaneous
backside events.

The solar source of the 2000 November 8 event (No.~92 in
Table~\ref{tab:table1}) is ascribed to an M7.8/1N flare in active
region (AR) 9212 \citep{Kurt2004} or AR~9213
\citep{NittaCliverTylka2003}. A related CME had an average speed
of 1738~km~s$^{-1}$ and extrapolated onset time of
$\approx\,$22:48 at $1R_{\odot}$ (CME catalog). A type II burst
started by 22:50 (HiRAS spectrometer), close to the CME onset, but
well before a weak microwave burst, which started after 23:04.
According to \citet{Zhang2001} and
\authorcite{Temmer08} (\yearcite{Temmer08,Temmer10}), the microwave
burst is expected $\sim 15$ min earlier.
\citet{NittaCliverTylka2003} proposed `\textit{the associated SEP
event appears to have originated in the non-active region eruption
rather than the M7.7 flare, although the entire complex of minor
active regions was probably involved at some level}'. However,
such a fast CME is not expected from a non-active region eruption.
A proton index $\delta_{p} = 1.64$ is atypically hard of SEP
events after such eruptions, in which usually $\delta_{p} > 2$
\citep{Chertok09}. These facts hint at a possible implication of a
backside eruption.

The 2001 December 26 event (No.~93) looks also strange. The source
of the SEP event is ascribed to the M7.1/1B flare in AR~9742,
N08\,W54 \citep{Kurt2004,Cliver2006}. Identification of the solar
source is hampered by a gap in EIT observations between 04:47 and
05:22. The situation on the Sun with several minor active regions
and post-eruption manifestations look similar to the preceding
event. The major role of a non-active region eruption for the SEP
event with $\delta_{p} = 1.17$ responsible for GLE63 looks still
more doubtful than for the 2000 November 8 event. The CME
(1446~km~s$^{-1}$) and the main microwave burst started at about
05:06, while the type II burst started by 04:50 (HiRAS). The soft
X-ray emission in this event rose much longer than in all other
GLE events of the solar cycle 23 \citep{Aschwanden2012}, while the
rise phase roughly displays the CME velocity \citep{Zhang2001}. A
strong shock wave is not expected from a gradually expanding CME.
Possibly, the M7.1 flare was preceded by a stronger backside
event.

The microwave burst in the 2002 April 21 event (No.~94) looks too
moderate in comparison with the big SEP event, the X1.2 flare, and
the fast CME (2393~km~s$^{-1}$). The near-limb location of the
flare site (S14~W84) implies complications like partial
occultation of the flaring region or some kind of absorption of
the microwave emission suggested by its atypically flat spectrum.
The presence of three major peaks in the microwave time profile,
first of which corresponded to the CME onset time, suggests more
than one eruption and additional features of this event such as
merger of two to three shock waves into a stronger one.

The 2012 May 17 event (No. 95) was responsible for GLE71. The M5.1
GOES importance was the lowest one among all GLEs of the solar
cycles 22--24 \citep{Cliver2006}. The CME had an average speed of
1582~km~s$^{-1}$. A flare ribbon crossed the sunspot umbra.
STEREO-A did not show any candidate for a stronger event behind
the west limb. Some kind of absorption of the microwave emission
is not excluded (the spectrum was also flat), but the moderate
GOES importance does not support underestimation of the flare
emission. Like the preceding event, the microwave time profile had
three peaks, first of which corresponded to the CME onset. The
moderate GOES importance and the microwave flux suggest a
possibility of escape of an unusually large fraction of
accelerated particles, including electrons, into the
interplanetary space. Analysis of this event is anticipated.

Thus, possible causes of the mM exceptions can be different. These
events deserve attention and further investigation. The group of
such events can be actually larger: we did not analyze SEP events
with $J_{100}< 10$ pfu or those occurring beyond the observational
daytime in Nobeyama.

\subsection{Account of Properties of Events}

Section~\ref{s-outcome} and Figure~\ref{fig:durations} confirm the
well-known fact of a poorer proton production of short-duration
events. The conclusion about the distinction of SEP events into
the `gradual' and `impulsive' categories (with an intermediate
`mixed' group) has been drawn from observations, which were
related to solar sources indirectly. The duration criterion
referring to microwave or soft X-ray bursts (see, e.g.,
\cite{Cliver1989}) also considers indirect outer manifestations of
solar events.

As shown in section~\ref{s-introduction}, the scenario of a solar
event offers very different opportunities for the development of a
shock wave and escape of protons from an active region. In a
typical powerful CME-productive event, a sharp eruption excites a
shock wave capable of particle acceleration, and the CME lift-off
makes possible escape of most flare-accelerated particles trapped
in the CME's flux rope. An opposite extremity of a confined flare
does not produce a shock wave, while escape is possible for those
accelerated particles, which are brought by drifts and diffusion
to open magnetic fields permanently existing in active regions.
The bulk of accelerated protons remains trapped. By chance, we
know that events No. 87 and 88 in September 2005 were most likely
confined flares. Their durations were short indeed. On the other
hand, the durations of the four mM exceptions were considerable
(Figure~\ref{fig:durations}).

Probably, failed eruptions like the event presented by
\citet{Ji2003} also do not produce shock waves, neither favor
particle escape. There may be confined eruptions, which excite
shock waves, but do not produce CMEs. Jet-like eruptions seem to
favor escape of flare-accelerated particles, while related shock
waves probably rather rapidly dampen for the absence of
significant energy supply from a trailing ejecta. The variety of
types of flares and eruptions might respond in their SEP outcome.
The most apparent, well-known indication of a shock wave, is a
type II burst (although the absence of a type II emission does not
guarantee the absence of a shock), and an indication of opening
magnetic configuration is a CME. We have found information about
CMEs for 52 events, which occurred since 1996 (after the launch of
SOHO) through 2012. It is presented by the qualifier Q0, Q1, Q2 as
a superscript in column~1 of Table~\ref{tab:table1} described in
section~\ref{s-data_table}. These 52 events are plotted in
Figure~\ref{fig:event_distribution_diag}. East events are
presented with the longitudinal correction. The mM exceptions are
qualified as Q2. The backside mO events are not shown for the
absence of any relevant information.
Figure~\ref{fig:event_distribution_diag} shows the following:

 \begin{figure}
  \begin{center}
    \FigureFile(85mm,60mm){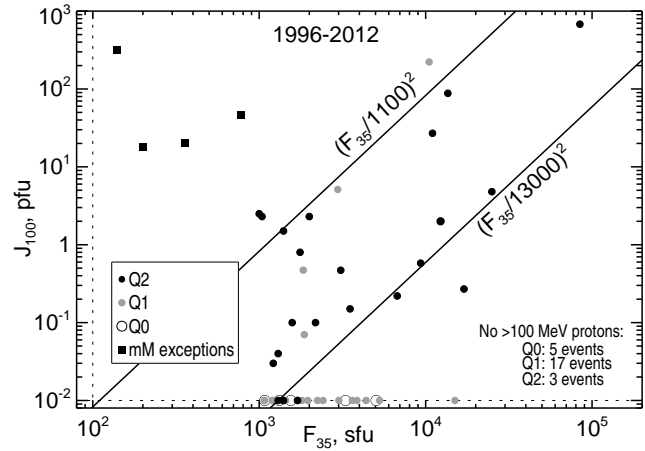}
  \end{center}
  \caption{Distribution of NoRP events, which occurred in 1996--2012,
in the same format as Figure~\ref{fig:event_distribution}, i.e.,
fluxes of $> 100$ MeV protons vs. radio fluxes at 35 GHz. The
symbols denote the qualifiers of the events according to the legend
in the frame. The black solid lines delimit the `main sequence'.}
 \label{fig:event_distribution_diag}
 \end{figure}

\begin{enumerate}

 \item

The majority of west events with both CMEs and type II bursts (Q2
events) belongs to the `main sequence'. Three
non-proton-productive Q2 events are nevertheless located close to
the `main sequence'.

 \item

No near-Earth fluxes of $> 100$~MeV protons were detected after
east events, which were associated with neither CME nor type II
burst. These are five Q0 events.

 \item

Most remaining events (17 Q1 events) did not produce near-Earth
protons with energies $> 100$~MeV. Four Q1 events belong to the
main sequence.

\end{enumerate}

A simple criterion follows from this distribution: if an event has
a west location and produces both CME and type II burst, then the
near-Earth flux of $> 10$~MeV protons expected from this event
should be most likely between $(F_{35}/1100)^2$ and
$(F_{35}/13000)^2$~pfu. This criterion can be possibly used as a
tentative basis for a future prompt diagnostics of proton events.

\section{Conclusion}
 \label{s-conclusion}

Recent observations and their studies have revealed still closer
relations between solar eruptions, flares, shock waves, and CMEs,
than previously assumed. This circumstance provides a basis to
expect a closer correspondence between parameters of near-Earth
proton enhancements and microwave bursts than the `Big Flare
Syndrome' can predict. This expectation has been mainly confirmed
in the present brief analysis of relations between about one
hundred of strong microwave bursts recorded in 1990--2012 with
NoRP at 35~GHz and near-Earth proton fluxes.

There is a scattered general correspondence between the peak flux
density at 35 GHz and peak flux of $>100$~MeV protons, $J_{100}
\approx (F_{35}/3800)^2$. In accordance with well-known patterns,
events with $F_{35} < 10^4$~sfu in far east active regions as well
as events with short-duration bursts have a reduced proton outcome
up to zero. On the other hand, most west events associated with
intense 35~GHz bursts, CMEs, and type II bursts produce the
near-Earth fluxes of $> 10$~MeV protons between $(F_{35}/1100)^2$
and $(F_{35}/13000)^2$~pfu. Overall, extreme long-duration bursts
at 35 GHz ($F_{35} > 10^4$~sfu) indicate large proton enhancements
with predominantly hard energy spectra, including GLEs. Large SEP
events are possible even with an eastern location of a solar
source region, if the 35 GHz burst is especially intense.

A morphological manifestation of a high-intensity burst at 35 GHz
is a flare occurring above the sunspot umbra. This is a different
indication of a possible SEP event. Thus, strong high-frequency
bursts or/and flare ribbons crossing the sunspot umbrae can be
employed to promptest alert of SEP events. However, in the case of
a backside flare, its microwave emission is occulted for
radiometers at Earth, while energetic particles propagating along
the Parker spiral can reach Earth.

Another limitation of diagnostic opportunities of microwave
observations is offered by challenging big SEP enhancements, which
rather rarely occur in association with moderate microwave bursts.
Such events deserve special attention and need investigating. Case
studies of various events can significantly contribute to better
understanding their SEP productivity and related conditions.

While some questions remain unanswered, it is clear that NoRP and
NoRH observations are highly important in further investigating
into the SEP problem.

\bigskip

Acknowledgements. We thank V. Kurt, A. Belov, A. Uralov, H.
Nakajima, B. Yushkov, K.-L. Klein, A. Tylka, S. White, Y. Kubo, N.
Nitta, and S. Kalashnikov for fruitful discussions and assistance.
We thank an anonymous referee for useful suggestions. We are
grateful to the instrumental teams operating the Nobeyama solar
facilities and GOES satellites. Data on CMEs have been taken from
the on-line CME catalog generated and maintained at the CDAW Data
Center by NASA and the Catholic University of America in
cooperation with the Naval Research Laboratory. SOHO is a project
of international cooperation between ESA and NASA.

This study was supported by the Russian Foundation of Basic
Research under grants 11-02-00757 and 12-02-00037, the Program of
the RAS Presidium No.~22, and the Russian Ministry of Education
and Science under State Contracts 16.518.11.7065 and
02.740.11.0576. N.M. was sponsored by a Marie Curie International
Research Staff Exchange Scheme Fellowship within the 7th European
Community Framework Programme.

\end{document}